\documentclass[12pt]{article}

\usepackage{amsmath}
\usepackage{amssymb}

\newcommand{\bel}[1]{\begin{equation}\label{#1}}
\newcommand{\ee}{\end{equation}}

\newcommand{\bal}[1]{\begin{eqnarray}\label{#1}}
\newcommand{\ea}{\end{eqnarray}}

\begin{document}

\title{A self-consistent quantal treatment of decay rates \\
       within the Perturbed Static Path Approximation }

\author{C. Rummel\thanks{e-mail: crummel@ph.tum.de} \,and
        H. Hofmann\thanks{e-mail: hhofmann@ph.tum.de} \\
\small\it{Physik-Department der TU M\"unchen, D-85747 Garching,
Germany}}
 \maketitle

\abstract{The framework of the Perturbed Static Path Approximation
(PSPA) is used to calculate the partition function of a finite
Fermi system from a Hamiltonian with a separable two body
interaction. Therein, the collective degree of freedom is
introduced in self-consistent fashion through a
Hubbard-Stratonovich transformation. In this way all transport
coefficients which dominate the decay of a meta-stable system are
defined and calculated microscopically. Otherwise the same formalism
is applied as in the Caldeira-Leggett model to deduce the decay rate
from the free energy above the so called crossover temperature $T_0$. \\
PACS number(s): 05.60.Gg, 82.20.Xr, 05.30.-d, 21.60.-n}

\tableofcontents

\section{Introduction}
\label{intro}

Commonly quantum versions of the decay rate of damped meta-stable
systems are treated on the basis of the Caldeira-Leggett model
\cite{ca.le.83}. For an exhaustive overview on this topic we may
refer to the review articles by P.~H\"anggi et al.
\cite{ha.ta.bo.90}, G.-L. Ingold in \cite{in.98} or the text books
by U.~Weiss \cite{weiss}. Unfortunately, this model assumes some
simplified coupling to a linear heat bath. Furthermore it does not
make any predictions about the input of such important quantities
as the potential energy and the inertia, which need to be chosen
on an entire phenomenological level. These features do not allow
the model to be applied to self-bound Fermi systems. There, one
would like to see the collective variables introduced in some
self-consistent fashion, with a microscopic treatment of {\em all}
transport coefficients alike.

One possible attempt to overcome these deficiencies is formulated
in \cite{hofrep} (with references to earlier papers) in connection
to nuclear physics. It is based on a quantal transport equation
which is derived within a locally harmonic approximation
exploiting linear response theory. It is this approximation which
allows one to treat a more complicated coupling between the
collective variable and the intrinsic degrees of freedom. A
transport equation necessarily describes evolution in real time.
Therefore, in barrier regions quantum effects can be accounted for
only above a critical temperature $T_c$, which is larger than the
so called crossover temperature $T_0$ which one encounters for
imaginary time propagation \cite{hh.in.th.93}. As one knows, the
same feature holds true also for the Caldeira-Leggett model
\cite{an.gr.in.95, an.gr.96}. Another disadvantage of the
derivation of this transport equation mentioned is that it bases
on the deformed shell model. Surely, it allows one to calculate
all transport coefficients on the same footing. But as one does
not start from a genuine two body interaction, self-consistency is
handled on a semi-microscopic level only.

It is the aim of the present paper to do first steps to overcome
these deficiencies. This is possible by adapting a previously
developed formalism to evaluate the partition function for bound
systems with separable two body interactions. One starts from the
so called Static Path Approximation (SPA) functional integrals as
an approximation to the classical or high temperature limit
\cite{al.zi.84, ar.be.la.pu.88, la.pu.bo.bro.90}. Then small scale
fluctuations around this static path are treated to second order,
such that quantum effects come in through local RPA. In the
literature this is referred to as RPA-SPA in \cite{pu.bo.bro.91},
the Perturbed Static Path Approximation (PSPA) in \cite{at.al.97}
(the name which we are taking over) or Correlated Static Path
Approximation (CSPA) in \cite{ro.ca.97}.

\section{Partition function of a finite Fermi system}
\label{Zustsum}

Finally, we are interested to generalize the formulas of
dissipative tunneling to a system where the collective degrees of
freedom are introduced self-consistently.  The simplest
Hamiltonian which may serve this purpose is of the following
structure
\bel{twobodham}
\hat{H} = \hat{H}_{0} + \frac{k}{2} \,\hat{F}\hat{F}\ ,
\end{equation}
with (hermitian) one body operators $\hat{H}_{0}$ and $\hat{F}$.
The product $\hat{F}\hat{F}$ mimics an effective separable two
body interaction. For isoscalar modes, the case we have in mind
predominantly, the coupling constant $k$ is negative \cite{bm}. As
we shall see later, the $\hat{F}$ describes one collective degree
of freedom. The ansatz (\ref{twobodham}) should be considered to
define a microscopic model for just this collective mode we want
to address to \cite{bm}. Neglecting spin and isospin degrees of
freedom, a general two body interaction may be written as a sum of
separable terms \bel{2bHamgen} \hat{H} = \hat{H}_{0} + \frac{1}{2}
\sum_{i} k_{i} \,\hat{F}_{i}\hat{F}_{i}\,.
\end{equation}
For instance, one might exploit an expansion into multipole
operators. In case the latter are not hermitian the product must
be replaced by $\hat{F}^\dagger_{i}\hat{F}_{i}$ (see e.g. section
4.4.7 of \cite{ring.schuck}).

\subsection{The general form of the partition function}
\label{genform}

The partition function of the grand canonical ensemble reads
\bel{Z}
Z(\beta) = \textrm{Tr} \,\exp \left( -\beta (\hat{H} - \mu \hat{A})
\right) = \textrm{Tr} \ \hat{U} \ ,
\end{equation}
where $\beta = 1/T$ is the inverse temperature and $\hat{H}$ is
the Hamiltonian (\ref{twobodham}). The chemical potential $\mu$ is
needed in order to keep the particle number
$\langle\hat{A}\rangle$ fixed on average. It would be more
appropriate to work with truly fixed particle number. But as we
are mainly interested in the dependence of transport properties on
excitation energy or temperature in this paper, this
simplification should be accepted. An exact evaluation of
(\ref{Z}) is prohibited by the presence of the two body
interaction. Treating the latter in mean field approximation
facilitates the calculation greatly. A convenient technical tool
to incorporate this approximation is to use functional integrals
(in imaginary time propagation) \cite{ne.or.88}, with which
fluctuations about the mean field may be treated as well. An
elegant form of handling this problem is given through the
Hubbard-Stratonovich transformation \cite{st.hu.5759}, by which
the collective variable $q(\tau)$ is introduced. To keep the
present exposition as short as possible we safe ourselves from
repeating the derivations of \cite{al.zi.84} - \cite{ro.ca.97} but
simply state the basic results which will then serve as the
starting point for our generalizations. Mind, however, that the
notation has been adapted to that used in transport theory
\cite{nat-adapt}.

After introducing the Fourier expansion of the collective variable
\bel{fluctcoord}
q(\tau) = q_{0} + \sum_{r \ne 0} q_{r} \,\exp (i\nu_{r}\tau) \ ,
\end{equation}
where the so called Matsubara frequencies
\bel{Matsubara} \nu_{r} = \frac{2\pi}{\hbar\beta} \,r \equiv
\frac{2\pi}{\hbar} \,r T \qquad \text{with} \qquad r=\pm 1,~\pm 2,
~\pm 3 ~\dots
\end{equation}
(in units with $k_{\textrm{B}}=1$) have been used,
the partition function may be written in the
following form within the PSPA (see eq.(21) of \cite{at.al.97}):
\bel{Z-aa}
Z(\beta) = \sqrt{\frac{\beta}{-2\pi k}}
\int_{-\infty}^{+\infty} \!\! dq_{0} \ e^{\frac{\beta}{2k}
q_{0}^{2}} \ z(\beta, q_{0}) \ \mathcal{C}(\beta, q_{0})\ .
\end{equation}
Here
\bel{zbeta}
z(\beta, q_{0}) = \textrm{Tr} \,\exp \left( -\beta (\hat{h}_{0}(q_{0})
- \mu \hat{A}) \right) = \prod_{l} \left( 1 + \exp \left( -\beta
(\epsilon_{l}(q_{0}) - \mu) \right) \right)
\end{equation}
is the grand canonical partition function belonging to the static part of
the Hamiltonian in mean field approximation
\bel{1bHam-stat} \hat{h}_{0}(q_{0}) = \hat{H}_{0} + \hat{F} q_{0}
\end{equation}
which is simply a sum over one body operators. The corresponding
one body Schr\"oding\-er equation at some given $q_{0}$ reads:
\bel{h0Schr}
\hat{h}_{0}(q_{0}) \,|l(q_{0})\rangle = \epsilon_{l}(q_{0}) \,|l(q_{0})\rangle
\end{equation}
The appearance of the $q_0$ reflects the static version of the
self consistency relation for the mean field,
\bel{selfmeanF} q= k \langle \hat F \rangle\ ,
\end{equation}
which relates the collective variable $q$ to the expectation value
of the operator $\hat F$. The exponent in the first factor of the
integrand of (\ref{Z-aa}) is easily understood as to represent the
static part of the correction $-k \langle\hat{F}\rangle^{2} / 2 $
which the energy $\langle \hat h_0 \rangle$ in the independent
particle picture gets from the two body interaction. Neglecting
the factor $\mathcal{C}$ one obtains the partition function in
{\em Static Path Approximation} (SPA) \cite{al.zi.84,
ar.be.la.pu.88, la.pu.bo.bro.90}
\bel{Z-SPA} Z^{\textrm{SPA}}(\beta) =
\sqrt{\frac{\beta}{-2\pi k}} \int_{-\infty}^{+\infty} \! dq_{0} \
e^{\frac{\beta}{2k} q_{0}^{2}} \ z(\beta, q_{0}) \equiv
\sqrt{\frac{\beta}{-2\pi k}} \int_{-\infty}^{+\infty} \! dq_{0} \
e^{-\beta{\mathcal{F}}^{\textrm{SPA}}(\beta, q_0)}\ .
\end{equation}
On the very right of (\ref{Z-SPA}) the symbol
${\mathcal{F}}^{\textrm{SPA}}(\beta, q_{0})$ has been introduced
to represent a free energy. It is not the one of the total system
(or total nucleus) which would be given by the relation
\bel{relZFREEen}
Z(\beta)= \exp \left( -\beta{\mathcal{F}}(\beta) \right) \ ,
\end{equation}
when the partition function is identified as
$Z(\beta) = Z^{\textrm{SPA}}(\beta)$. Rather, the
${\mathcal{F}}^{\textrm{SPA}}(\beta, q_0)$ represents the free
energy of the system of nucleons whose mean field is kept fixed at
the $q_0$. In a common language of transport theory one would call
it the free energy of the intrinsic degrees of freedom.

So far any contribution from the dynamics in the collective
variable $q(\tau)$ has been neglected. Formally this may be
accounted for by writing the correction factor as the following
path integral \cite{at.al.97}
\bel{corrfactor} \mathcal{C}(\beta, q_{0}) = \int {\cal D}'q \ \exp
\left( \frac{\beta}{k} \sum_{r>0} |q_{r}|^{2} + \textrm{ln} \langle
\hat{\mathcal{U}}_{q} \rangle_{q_{0}} \right)
\end{equation}
with the measure
\bel{mass}
{\cal D}'q = \lim_{\substack{N\rightarrow\infty \\
N\varepsilon = \hbar\beta}} \prod_{r=1}^{(N-1)/2}
\frac{\beta}{-\pi k} \ d\textrm{Re}(q_{r}) \,d\textrm{Im}(q_{r})
\end{equation}
In (\ref{corrfactor}) there appears the thermal expectation value
of an evolution operator $\hat{\mathcal{U}}_{q}$ which can be
expressed by the following (imaginary) time-ordered product
\bel{meanUq} \langle \hat{\mathcal{U}}_{q} \rangle_{q_{0}} =
\frac{1}{z(\beta, q_{0})} \ \textrm{Tr} \left( \exp (-\beta
\hat{h}_{0}(q_{0})) \ \hat{\cal T} \exp \left[ -\frac{1}{\hbar}
\int_{0}^{\hbar\beta} d\tau \,\hat{h}_{1}(\tau, q_{r}) \right]
\right) \ .
\end{equation}
The Hamiltonian
\bel{1bHam-td-int}
\hat{h}_{1}(\tau, q_{r}) = \hat{F}(\tau) \,\delta q(\tau)
\end{equation}
may be understood as the time dependent correction to the static
mean field given in (\ref{1bHam-stat}). Here, the time-dependence
of the operator part is defined as
\bel{interact}
\hat{F}(\tau) = e^{\hat{h}_{0}(q_{0}) \tau/\hbar} \ \hat{F}
\ e^{-\hat{h}_{0}(q_{0}) \tau/\hbar} \ ,
\end{equation}
which means through the interaction picture based on the
Hamiltonian $\hat h_0(q_0)$ of (\ref{1bHam-stat}) and thus depends
on the static $q_{0}$. The fluctuation of the collective variable
$\delta q(\tau) = q(\tau) - q_{0}$ in (\ref{1bHam-td-int}) is related
to the fluctuating mean field through (\ref{selfmeanF}). It may be noted
that the $\tau$-dependence of the c-number $q$ is meant to be the correct
one, not that of any interaction picture.

The partition function (\ref{Z-aa}) may finally be written in the
following compact form
\bel{Z-ab} Z(\beta) = \sqrt{\frac{\beta}{-2\pi k}}
\int_{-\infty}^{+\infty} \!\! dq_{0} \ \exp (-\beta {\mathcal{F}}(\beta, q_{0}))
\end{equation}
if again one uses the concept of the "intrinsic free energy" which
now is given by
\bel{F-a}
{\mathcal{F}}(\beta, q_{0}) = -\frac{1}{2k} \,q_{0}^{2}
- \frac{1}{\beta} \ \textrm{ln} \,z(\beta, q_{0})
- \frac{1}{\beta} \ \textrm{ln} \,\mathcal{C}(\beta, q_{0}) \ .
\end{equation}

\subsection{The Perturbed Static Path Approximation (PSPA)}
\label{PSPA}

We are now going to evaluate the general formula
(\ref{corrfactor}) within the so called PSPA. It is defined as
that approximation in which the exponent appearing in
(\ref{meanUq}) is  expanded to second order in the $q_r$. This
leads to the common Gaussian approximation which is known to be
related to the semi-classical limit. Following \cite{at.al.97} one
may write
\bel{logU} \textrm{ln}
\,\langle\hat{\mathcal{U}}_{q}\rangle_{q_{0}}^{\textrm{PSPA}} =
\frac{1}{2\hbar^{2}} \sum_{r,s\not=0} q_{r}q_{s}
\int_{0}^{\hbar\beta} \!\!d\tau \int_{0}^{\hbar\beta} \!\!d\sigma
\,e^{i\nu_{r}\tau} e^{i\nu_{s}\sigma} \langle \hat{\cal T}
\hat{F}(\tau)\hat{F}(\sigma) \rangle_{q_{0}} \ ,
\end{equation}
with the $\tau$-dependence of the operators as defined in
(\ref{interact}). Likewise, according to (\ref{meanUq}), the
expectation value is to be calculated with the density operator
corresponding to the same unperturbed Hamiltonian
$\hat{h}_{0}(q_{0})$. It is this feature which will allow us to
introduce and work with response functions. As we shall see below,
this is of advantage for  at least two reasons, which in a sense
are related to each other. The final result, say for the decay
rate of metastable states, has much in common with the linear
response formulation of transport theory within a locally harmonic
approximation \cite{hofrep}. From this approach one knows how the
response functions have to be modified in order to introduce
dissipation.

\subsubsection{Exploiting Green and response functions}
\label{greenresp}

The time ordered average in (\ref{logU}) can be identified with
the two body Matsubara function of the one body operator $\hat{F}$
\cite{mahan}:
\bel{2bMat-t} \tilde{\cal G}(q_{0}, \tau - \sigma) =
-\frac{1}{\hbar} \,\langle \hat{\cal T}\hat{F}(\tau)\hat{F}(\sigma)
\rangle_{q_{0}}
\end{equation}
On the other hand the retarded FF-response function is given by
\bel{ret2bGreen-t}
\tilde{\chi}^{R}(q_{0}, t - s) =
\frac{i}{\hbar} \,\theta(t - s)
\,\langle\hat{F}(t)\hat{F}^{\dagger}(s) -
\hat{F}^{\dagger}(s)\hat{F}(t)\rangle_{q_{0}} =
\tilde{\chi}(q_{0}, t - s)\ .
\end{equation}
(Henceforth, we shall omit the upper index "R"). It describes the
response of the expectation value $\langle \hat{F} \rangle$ to the
variations of $q$ in real time evolution,
\bel{defResp} \langle \hat{F} \rangle_{q_{0}}(t) =
-\int_{-\infty}^{\infty} ds \ \chi(q_{0}, t - s) \,(q(s) -
q_{0})\,,
\end{equation}
The spectral representations ${\cal G}(q_{0}, i\nu_{r})$ of
(\ref{2bMat-t}) and $\chi(q_{0}, \omega)$ of (\ref{ret2bGreen-t})
are obtained by  Fourier series and Fourier transformations,
respectively. As both have the same spectral density, one may
prove \cite{mahan} them to be connected by the analytic
continuation
\bel{contin}
{\cal G}(q_{0}, i\nu_{r}) \quad
\overleftrightarrow{i\nu_{r} \leftrightarrow \omega + i\epsilon}
\quad -\chi^{R}(q_{0}, \omega) \ .
\end{equation}
The response function may be continued to the whole complex plane
via \cite{hofrep}
\bel{X}
X(q_{0}, z) = \int_{-\infty}^{\infty} \frac{d\Omega}{\pi}
\,\frac{\chi''(q_{0}, \Omega)}{\Omega - z} \qquad \textrm{for}
\qquad \textrm{Im} \,z \not= 0 \ ,
\end{equation}
with $\chi''(q_{0}, \omega)$ being the imaginary (dissipative)
part of $\chi(q_{0}, \omega)$. The form (\ref{X}) defines two
branches. The one which is analytic in the upper half plane
coincides with the retarded function $\chi^{R}(q_{0}, z)$ and the
one analytic in the lower half plane defines the advanced function
$\chi^{A}(q_{0}, z)$. Both branches may be continued analytically
into the other half planes. Below we will make use only of the
retarded response $ \chi^{R}(q_{0}, z) \equiv \chi(q_{0}, z)$. On
the imaginary axis ($z = iw$ with $w \in \mathbf{R}$) it has the
following symmetry properties:
\bel{Xsym} (\chi(q_{0}, iw))^{*} = \chi(q_{0}, (iw)^{*})
= \chi(q_{0}, -iw) = \chi(q_{0}, iw)
\end{equation}
This property, together with the relations (\ref{2bMat-t}) and
(\ref{contin}) may be exploited to calculate the $\tau$-integrals
in (\ref{logU}) as
\bel{logU2}
\textrm{ln} \,\langle\hat{\mathcal{U}}_{q}\rangle_{q_{0}}^{\textrm{PSPA}}
 = \beta \sum_{r>0} |q_{r}|^{2} \,\chi(q_{0}, i\nu_{r}) \,.
\end{equation}
Mind that because of the reality of the collective variable
one has $q_{r}^{*} = q_{-r}$. The result (\ref{logU2})
may be plugged into (\ref{corrfactor}) to arrive at the following
form
\bel{C-c} \mathcal{C}^{\textrm{PSPA}}(\beta, q_{0}) = \int {\cal
D}'q \ \exp \left( \frac{\beta}{k} \sum_{r>0} \left( 1 +
k\chi(q_{0}, i\nu_{r}) \right) |q_{r}|^{2} \right) \ .
\end{equation}
The remaining integrals hidden in ${\cal D}'q$ are of Gaussian type.
As we stick to the case $k < 0$, they cause no problem as long as
\bel{Konvbed}
1 + k\chi(q_{0}, i\nu_{r}) > 0 \qquad \textrm{for} \qquad r > 0 \ .
\end{equation}
As we shall see soon this leads to a condition on the temperature
below which the PSPA breaks down, as already noticed in
\cite{pu.bo.bro.91, at.al.97}. Here, this condition only has
been rewritten in terms of the response functions used in the
linear response approach to nuclear transport (see
e.g.\cite{hofrep}). In this language the final result for
$\mathcal{C}^{\textrm{PSPA}}$ reads
\bel{C}
\mathcal{C}^{\textrm{PSPA}}(\beta, q_{0}) = \prod_{r>0} \left( 1 + k\chi(q_{0},
i\nu_{r}) \right)^{-1}
\end{equation}
and that for the partition function of the PSPA becomes
\bal{ZustSum-a}
Z^{\textrm{PSPA}}(\beta) & = & \sqrt{\frac{\beta}{-2\pi k}} \int_{-\infty}
^{+\infty} \!\! dq_{0} \ e^{-\beta {\mathcal{F}}^{\textrm{SPA}}(\beta, q_{0})}
\ \mathcal{C}^{\textrm{PSPA}}(\beta, q_{0}) \nonumber \\
& = & \sqrt{\frac{\beta}{-2\pi k}} \int_{-\infty}^{+\infty} \!\! dq_{0}
\ e^{\frac{\beta}{2k} q_{0}^{2}} \ z(\beta, q_{0}) \ \prod_{r>0}
(1 + k\chi(q_{0}, i\nu_{r}))^{-1} \ .
\end{eqnarray}

\subsubsection{Response functions in the independent particle model}
\label{resp-ipm}

Before we are going to discuss further the condition
(\ref{Konvbed}) in the next subsection, let us recall how the
response function looks like in the model of independent
particles, as defined by the Hamiltonian $\hat h _0$ of
(\ref{1bHam-stat}). It is not difficult to convince oneself of the
following form for the dissipative part of the FF-response function
\bel{dResp}
\chi''(q_{0}, \omega) = -\frac{\pi}{\hbar}
\sum_{l,k} |F_{lk}(q_{0})|^{2} \ n_{lk}(q_{0})
\ \delta(\omega - \epsilon_{lk}(q_{0})/\hbar) \ ,
\end{equation}
where
\bel{dedifepoc}
\begin{split} F_{lk}(q_{0}) & = \langle
l(q_{0})| \,\hat{F} \,|k(q_{0}) \rangle \\ \epsilon_{lk}(q_{0}) &
= \epsilon_{l}(q_{0}) - \epsilon_{k}(q_{0}) \\ n_{lk}(q_{0}) & =
n(\epsilon_{l}(q_{0})) - n(\epsilon_{k}(q_{0}))
\end{split}
\end{equation}
and $n(e)$ being the Fermi occupation numbers
\bel{Fermi}
n(e) = \frac{1}{1 + \exp (\beta (e - \mu))} \ .
\end{equation}
Within this model it can easily be seen, that the $\chi(q_{0}, z)$ is given by
\bel{X-MuT} \chi(q_{0}, z) =
-\frac{1}{\hbar} \sum_{l,k} |F_{lk}(q_{0})|^{2}
\,\frac{n_{lk}(q_{0})}{\epsilon_{lk}(q_{0})/\hbar - z}
\end{equation}
Notice, please, that along the real axis the $z$ must be chosen
identical to $\omega + i\epsilon$. For details about these
response functions we may refer to \cite{hofrep}.

\subsubsection{The crossover temperature} \label{crossover}

Let us elaborate now on the convergence condition (\ref{Konvbed})
for the $q_{r}$-integrals in (\ref{C-c}), finally to establish
connection to an analogous condition which shows up when treating
dissipative tunneling at finite temperature within the Caldeira-Leggett model
\cite{la.ov.8384, gr.we.ha.84}. To this end the following identity
is useful \cite{at.al.97}
\bel{def-omnu}
1 + k\chi(q_{0}, i\nu_{r}) = \frac{\prod_{\nu}
(\nu_{r}^{2} + \omega_{\nu}^{2}(q_{0}))}{\prod_{k>l}' (\nu_{r}^{2}
+ (\epsilon_{lk}(q_{0})/\hbar)^{2})} \ ,
\end{equation}
which is valid for all $r \not= 0$. The frequencies
$\omega_{\nu}(q_{0})$ appearing here are those of the local RPA
associated to the local vibrations of the mean field around
$q_{0}$. They satisfy a secular equation \cite{at.al.97}, which
can easily be brought to the form
\bel{RPA} 1 + k \chi(q_{0}, z) = 0
\end{equation}
by analytically continuing the function $\mathcal{G}(q_{0},
i\nu_{r})$ to complex $z$ by way of (\ref{contin}) and (\ref{X}).
As the denominator of the ratio on the right of (\ref{def-omnu}) is
real and positive the condition (\ref{Konvbed}) can be
reformulated as
\bel{Konvbed2}
\prod_{\nu} \left( \nu_{r}^{2} + \omega_{\nu}^{2}(q_{0}) \right) > 0
\end{equation}
as already mentioned in \cite{pu.bo.bro.91, at.al.97}. In case
that all local RPA modes are stable, and hence that all
$\omega_{\nu}(q_{0})$ are real, the condition is fulfilled for any
temperature, viz for $T \geq T_{0} \equiv 0$ (mind
(\ref{Matsubara})). For {\em unstable RPA modes}, on the other
hand, {\em one} pair of corresponding frequencies
$\omega_{\nu}^{\text{inst}}(q_{0})$ becomes purely imaginary, in
which case (\ref{Konvbed2}) can be fulfilled only above a certain
minimal temperature $T_{0}(q_{0})$. The latter may vary with
$q_0$, but it is possible, of course, to define a minimal global
temperature $T_0$ by
\bel{T0-a}
T_{0} =  \textrm{max} \ \frac{\hbar|\omega_{\nu}^{\text{inst}}(q_{0})|}{2\pi}
\end{equation}
such that (\ref{Konvbed}) is fulfilled for all $T > T_0$.  This
temperature is identical \cite{khqstsu.F} to the so called
"crossover temperature" (here of course for an undamped system)
that shows up in the Caldeira-Leggett model when dealing with
unstable modes of dissipative quantum systems \cite{gr.we.ha.84}.
There, the notion "crossover" indicates a transition
in the nature of the decay of a metastable system. Above $T_0$ the
process is dominated by thermally activated decay ("thermal
hopping") with the effects of genuine barrier penetration in the
quantum sense to become dominant only below this $T_0$ (called
"dissipative tunneling" for damped quantum systems). Evidently, in
a typical situation, the $T_0$ of (\ref{T0-a}) would correspond to
that $q_0$ where the top of the barrier is located.

\section{The PSPA for dissipative phenomena}
\label{Diss}

To elaborate on the connection to the treatment of dissipative
tunneling within the Caldeira-Leggett model we need to introduce
dissipation. As mentioned previously, the most natural way is
through the response function. This can best be seen at the
secular equation (\ref{RPA}). For real $z=\omega=\omega^*$, for
which the response function splits into its real (reactive) and
imaginary (dissipative) part, $\chi(\omega)  = \chi^\prime(\omega)
+ i \chi^{\prime\prime}(\omega)$, one gets \bel{sec-u-d}
  \begin{split}
    1 + k \chi^\prime(\omega)  & = 0 \\
    \chi^{\prime \prime}(\omega)  & = 0 \ .
  \end{split}
  \end{equation}
Whenever the function  $\chi^{\prime \prime}(\omega)$ is given by
a discrete sum of $\delta$-functions located at
$\epsilon_{lk}(q_{0})/\hbar$, as shown in (\ref{dResp}), the
second equation is automatically fulfilled at the solutions
$\omega_\nu$  of the first equation. These solutions are either
real or purely imaginary without any sign of dissipation,
reflecting the fact that the local RPA as discussed above
corresponds to time reversible dynamics. This argument shows that
irreversibility is intimately related to the functional form of
the dissipative part $\chi^{\prime \prime}(\omega)$ of the
response function. A genuinely microscopic approach would require
to consider explicitly couplings of the simple particle-hole
configurations to more complicated states \cite{ne.or.88}.
Definitely, this is beyond the scope of the functional integral
method underlying the present model. In a more phenomenological
approach one might argue to dress the single particle states with
complex self-energies which itself may vary with temperature, for
details see \cite{hofrep}  or \cite{iv.hh.99} where the inclusion
of pairing is discussed. An even simpler way is to effectively
perform the transition to a continuous spectrum, which directly
corresponds to the procedure one employs in the Caldeira-Leggett
model in typical solid state applications \cite{ca.le.83} -
\cite{weiss}, \cite{gr.we.ha.84}. However, even for a finite
nucleus such a transition is justified for not too small
excitations. Indeed, as one knows from nuclear reaction theory
\cite{feshbach}, for not too small energies resonances do overlap,
implying that the true compound states lie dense for excitations
above about $10-20$ MeV. On the level of the independent particle
model one simply might employ energy averages, which in turn are
related to finite observation times of the system; for details the
reader may be referred to \cite{hofrep}.

In this paper we would not like to penetrate any further into this
discussion. Rather, in the sequel we would like to assume the
$\chi^{\prime \prime}(\omega)$ to be a continuous function of
$\omega$. In this case the secular equation (\ref{RPA}) may no
longer be written as in (\ref{sec-u-d}) and its solutions become
complex quantities. To be specific, instead of (\ref{dResp}) we
like to suggest and work with the following model function
consisting of two Lorentzians of width $\Gamma(q_{0})$:
\bel{chipp-Lorentz}
\chi''(q_{0},\omega) = F^{2}(q_{0}) \left( \frac{\Gamma(q_{0})/2}
{(\omega - {\cal E}(q_{0}))^{2} + (\Gamma(q_{0})/2)^{2}}
- \left( {\cal E} \leftrightarrow -{\cal E} \right) \right)
\end{equation}
It may be characterized as a generalization of the degenerate
model often used in nuclear physics (see e.g. \cite{bm}) to one
where the nucleonic states are spread over a certain region
determined by the width $\Gamma(q_{0})$. The strength of these
intrinsic excitations is parameterized by the quantity
$F^{2}(q_{0})$. A straight forward generalization could be seen in
a summation of more than one term. In a sense the reduced form
(\ref{chipp-Lorentz}) corresponds to what has been called the "one
pole approximation" (see e.g. \cite{hofrep}). It is valid whenever
the strength distribution is dominated by one peak, which then
finally implies to have one prevailing collective mode. The
parameters appearing in (\ref{chipp-Lorentz}) could be calculated
in various ways, as indicated within the linear response approach,
for instance, but even the Random Matrix Model (RMM) might be used
(see e.g. \cite{hofrep}).

Inserting the spectral density (\ref{chipp-Lorentz}) into
(\ref{X}) the full response function can be calculated, which is
needed both for the secular equation (\ref{RPA}) as well as for
the condition (\ref{Konvbed}). The integral can be carried out
with the help of the residue theorem noticing that the integrand
has five poles altogether, situated at $\Omega = z$ and $\Omega =
\pm {\cal E}(q_{0}) \pm i\Gamma(q_{0})/2$, and closing the loop in
the appropriate half plane. The final result for the retarded response
function reads
\bel{Xbar+}
\chi(q_{0}, z) = F^{2}(q_{0}) \ \frac{{\cal
E}(q_{0})}{{\cal E}(q_{0})^{2} + (\Gamma(q_{0})/2)^{2} -
i\Gamma(q_{0}) \,z - z^{2}}\ ;
\end{equation}
(The advanced response function would be obtained by changing
$-i$ into $+i$). For the condition (\ref{Konvbed}) one needs to
know this function along the positive imaginary axis. There, the
denominator is always positive implying that $\chi(q_{0}, iw)$ is
finite for real $w$. Furthermore, it is seen that $\chi(q_{0},
iw)$ still is real for continuous spectra.

\subsection{Transport coefficients of collective motion}
\label{coll}

We are now going to write the secular equation for collective
motion in terms of transport coefficients, as it is known for the
damped oscillator. This is achieved best by rewriting
(\ref{Xbar+}) in the form of the oscillator response function
\bel{OSZ} \chi(q_{0},z) = \frac{-1}{M(q_{0})} \ \frac{1}{z^{2} +
i\Gamma(q_{0}) z - \Omega^{2}(q_{0})} \equiv
\chi_{\textrm{osc}}(q_{0},z) \ .
\end{equation}
The  parameters introduced here correspond to the {\em nucleonic}
(or "intrinsic") motion at any value of $q_{0}$ and are {\em uniquely}
derived from (\ref{Xbar+}) as follows:
\begin{eqnarray}
M(q_{0}) & = & -\frac{1}{2} \,\left. \frac{\partial^{2} \chi^{-1}(q_{0}, z)}
{\partial z^{2}} \right|_{z=0} = \frac{1}{{\cal E}(q_{0}) \ F^{2}(q_{0})}
\label{inertia} \\
M(q_{0}) \,\Omega^{2}(q_{0}) & = & \chi^{-1}(q_{0},z=0) \quad\qquad =
\frac{{\cal E}(q_{0})^{2} + (\Gamma(q_{0})/2)^{2}}
{{\cal E}(q_{0}) \ F^{2}(q_{0})}
\label{stiffness} \\
M(q_{0}) \,\Gamma(q_{0}) & = & i \,\left. \frac{\partial \chi^{-1}(q_{0}, z)}
{\partial z} \right|_{z=0} \ \quad = \frac{\Gamma(q_{0})} {{\cal E}(q_{0})
\ F^{2}(q_{0})}
\label{damping}
\end{eqnarray}
These transport coefficients may be interpreted as the (local)
coefficients of inertia, frequency and friction for the nucleonic
mode. Plugging (\ref{OSZ}) into (\ref{RPA}) one obtains
\bel{sekular-a}
0  =  1 + k \,\chi(q_{0}, z) = \frac{z^{2} +
i\Gamma(q_{0})z - \Omega^{2}(q_{0}) - k/M(q_{0})} {z^{2} +
i\Gamma(q_{0})z - \Omega^{2}(q_{0})} \ .
\end{equation}
This equation may be fulfilled only for a vanishing numerator,
which leads to the secular equation for the local frequencies
$z^{\pm}(q_{0})$ of collective motion, namely
\bel{sekular-b}
\left( z^{\pm} \right)^{2} + i\Gamma(q_{0}) z^{\pm} - \varpi^{2}(q_{0}) = 0 \ ,
\end{equation}
with the local {\em collective} frequency being defined as
\bel{freqcoll}
\varpi^{2}(q_{0}) = \Omega^{2}(q_{0}) + k / M(q_{0}) < \Omega^{2}(q_{0}) \ .
\end{equation}
The last inequality is given because we are dealing with isoscalar
modes where $k < 0$. Notice that the collective frequency $\varpi$ may become
purely imaginary, whereas the intrinsic one $\Omega$ is always
real (see (\ref{stiffness})).

Now the frequencies $z^{\pm}(q_{0})$ are no longer real quantities.
A convenient form is seen to be:
\bel{koll}
z^{\pm}(q_{0}) = |\varpi(q_{0})| \left( \pm \sqrt{\textrm{sgn}
\,\varpi^{2}(q_{0}) - \eta^{2}(q_{0})} - i\eta(q_{0}) \right) \ ,
\end{equation}
with
\bel{defvisc}
\eta(q_{0}) = \frac{\Gamma(q_{0})}{2 |\varpi(q_{0})|} \ .
\end{equation}
The dimensionless parameter $\eta(q_{0})$  measures the degree of
damping: It is smaller (larger) than $1$ if the (local) collective
motion is underdamped (overdamped). In the stable case
$\varpi^{2}(q_{0}) > 0$ the frequencies $z^{\pm}(q_{0})$ of
(\ref{koll}) are found in the lower complex half plane
symmetrically to the imaginary axis for $\eta(q_{0}) < 1$ and on
the negative imaginary axis for $\eta(q_{0}) > 1$. In the unstable
case $\varpi^{2}(q_{0}) < 0$ they always lie on the imaginary
axis, but now the frequency $z^{+} (q_{0})$ is in the upper half
plane.

It may be worth while to briefly compare (\ref{sekular-a}) with
the undamped case. It is easily recognized that for vanishing
$\Gamma$ the form (\ref{sekular-a}) turns into (\ref{RPA}) under the
following conditions:
(a) eq. (\ref{def-omnu}) is evaluated at $z$ instead of $i\nu_{r}$,
(b) simply one (pair of) local collective mode(s) $\varpi(q_{0})$ is considered
instead of all local RPA modes $\omega_{\nu}(q_{0})$,
(c) the intrinsic frequencies $\epsilon_{lk}(q_{0}) / \hbar$ are replaced
by $\Omega(q_{0})$.

\subsection{The crossover temperature for damped motion}
\label{T0glob}

There is no change in the condition (\ref{Konvbed}) for
convergence of the integrals in (\ref{C-c}). It is only that
(\ref{Konvbed}) takes on a different form in terms of the
transport coefficients. Moreover, the $\chi(q_{0},i\nu_{r})$ may be
expressed by the transport coefficients by making use of
(\ref{sekular-a}). In this way (\ref{Konvbed}) turns into
\bel{T0nur}
\nu_{r}^{2} + \Gamma(q_{0}) \,\nu_{r} + \varpi^{2}(q_{0}) > 0 \ ,
\end{equation}
as a natural generalization of (\ref{Konvbed2}). Still, for a real
collective frequency this condition is always fulfilled. For a
purely imaginary one, on the other hand, (\ref{T0nur}) can be
fulfilled only if the $\nu_{r}$ is larger than
\bel{nur+} \nu_{r}^{+} = |\varpi(q_{0})| \left( -\eta(q_{0}) +
\sqrt{\eta^{2}(q_{0}) - \textrm{sgn} \,\varpi^{2}(q_{0})}
\right) \ .
\end{equation}
Hence, $T$ has to be larger than the local crossover temperature
\bel{T0loc} T_{0}(q_{0}) = \frac{\hbar |\varpi(q_{0})|}{2\pi}
\,\left(\sqrt{1 + \eta^{2}(q_{0})} - \eta(q_{0}) \right)\ .
\end{equation}
Evidently, the $T_{0}(q_{0})$ is decreasing with growing damping strength
$\eta(q_{0})$. For $\eta(q_{0}) \gg 1$  one has $T_{0}(q_{0})
\sim 1/2\eta(q_{0})$. The global crossover temperature, finally,
has to be defined as
\bel{T0-b} T_{0} = \textrm{max} \,T_{0}(q_{0})\ .
\end{equation}
For vanishing damping we recover (\ref{T0-a}) with
$\varpi(q_{0})$ being identical to $\omega_{\nu}(q_{0})$.

\section{The fission rate within the PSPA} \label{decrate}

Imagine that we are given a heavy nucleus which may decay by
fission, a process which is to be understood as collective motion
across a barrier. It is known that at smaller temperatures this
barrier may have substructure due to shell effects. Such details
shall be neglected here. Rather we shall assume the process to be
dominated by just one potential minimum and one pronounced
barrier. Likewise, we shall discard any evaporation of light
particles and $\gamma$'s. Moreover, the transfer of energy from
the collective degree of freedom to the nucleonic ones will be
supposed not to change much the latter's temperature. Under such
circumstances the previously discussed path integral formulation
may be applied, with a fixed temperature. As noted earlier, for
the PSPA we expect great similarities to processes which are
described on the basis of the Caldeira-Leggett Hamiltonian.

There, the decay rate $R$ of unstable systems at not too small
temperatures is traced back to the imaginary part of the free
energy. As can be seen in the literature, see e.g. \cite{weiss},
\cite{in.98} or \cite{ha.ta.bo.90}, for  $T>T_{0}$ the following
formula is in wide use
\bel{-T>T0}
R = -\frac{2}{\hbar} \ \frac{T_{0}}{T} \ \textrm{Im}
\,{\mathcal{F}}(\beta)\,.
\end{equation}
It has originally been postulated in \cite{la.67, aff.81} and, in
strict sense, still lacks a general proof from first principles.
However, it can be said that it is capable of reproducing
correctly certain limits. For instance, one recovers correctly
Kramers' high temperature limit, and in the quantum limit one gets
the same functional form as obtained with real-time path integrals
\cite{an.gr.in.95} or in a quantum transport theory
\cite{hh.in.th.93}.

To evaluate the imaginary part of the free energy one still uses
the relation ${\mathcal{F}}(\beta) = -T \ \textrm{ln} \,Z(\beta)$
to the partition function. For an unstable system the latter
attains an (exponentially small) imaginary part. Following Langer
\cite{la.67} this may be shown by applying the saddle point
approximation and distorting the integration contour into the
complex plane at the barrier. Expanding the logarithm to first
order in the exponentially small quantity $\textrm{Im} \,Z(\beta)
/ \textrm{Re} \,Z(\beta)$ the imaginary part of the free energy
becomes:
\bel{-ImF}
\textrm{Im} \,{\mathcal{F}}(\beta) \approx
-T \ \frac{\textrm{Im} \,Z(\beta)}{\textrm{Re} \,Z(\beta)}
\end{equation}
Plugging (\ref{-ImF}) into (\ref{-T>T0}) we obtain
\bel{-ImFrate}
R = \frac{2T_{0}}{\hbar} \ \frac{\textrm{Im} \,Z(\beta)}{\textrm{Re}
\,Z(\beta)} \ .
\end{equation}
The partition functions appearing here may be evaluated within the
PSPA extending formula (\ref{ZustSum-a}) to a dissipative system
as outlined in section \ref{Diss}. Applying the saddle point
approximation to the $q_{0}$-integral in (\ref{ZustSum-a}) we
obtain
\bel{ZS-min}
\left. Z^{\textrm{PSPA}}(\beta) \right|_{q_{a}} =
\frac{1}{\sqrt{-k \,C_{\mathcal{F}}(q_{a})}}
\ \exp \left( -\beta
{\mathcal{F}}^{\textrm{SPA}}(\beta, q_{a}) \right) \
{\mathcal{C}}^{\textrm{PSPA}}(\beta, q_{a})
\end{equation}
as the contribution from the minimum and the purely imaginary
expression
\bel{ZS-barr}
\left. Z^{\textrm{PSPA}}(\beta) \right|_{q_{b}} =
\frac{i}{2\sqrt{-k \,|C_{\mathcal{F}}(q_{b})|}}
\ \exp \left(
-\beta{\mathcal{F}}^{\textrm{SPA}}(\beta, q_{b}) \right) \
{\mathcal{C}}^{\textrm{PSPA}}(\beta, q_{b})
\end{equation}
as the contribution from the barrier. Here, the stiffnesses
\bel{CFT-def}
\partial^{2} {\mathcal{F}}^{\textrm{SPA}}(\beta, q_{0}) /
\partial q_{0}^{2} = C_{\mathcal{F}}(q_{0})
\end{equation}
of the SPA free energy at fixed temperature appear, as it was assumed
that the integrand is dominated by the exponential and that the correction
factor $\mathcal{C}^{\textrm{PSPA}}(\beta, q_{0})$ varies smoothly
with $q_{0}$. The stationary points are thus defined by this free
energy through $\partial {\mathcal{F}}^{\textrm{SPA}} /
\partial q_{0} = 0$. Evaluating  the intrinsic
free energy in SPA from (\ref{F-a}) with ${\cal C} \equiv 1$ it is
easy to convince oneself that the extremal points fulfill the
relation
\bel{q0bar} q_{a/b} = k \langle\hat{F}\rangle_{q_{a/b}} \ .
\end{equation}
The derivatives of the eigenvalues $\epsilon_{l}(q_{0})$ with
respect to $q_{0}$ needed here may be obtained from
time-independent perturbation theory. In (\ref{q0bar}) the indices
$a$ and $b$ stand for the minimum and the maximum (or barrier) of
${\mathcal{F}}^ {\textrm{SPA}}(q_{0})$, respectively. The
relations (\ref{q0bar}) are nothing else but the self consistency
condition (\ref{selfmeanF}) applied to the two stationary points
of the system. Whereas (\ref{ZS-min}) was obtained through the
common Gaussian integrals of the saddle point approximation, for
(\ref{ZS-barr}) the integration contour had to be deformed such
that it runs parallel to the positive imaginary axis. This is the
reason for the additional factor 2 in the denominator of
(\ref{ZS-barr}), see \cite{col.78, aff.81, la.67}.

The generalization of the PSPA correction factor (\ref{C}) to
damped quantum systems may be performed by replacing the response
function (\ref{X-MuT}) of the independent particle model by its
continuous version (\ref{Xbar+}). Furthermore, we may make use of
the transport coefficients introduced in section \ref{coll}. In this
way one gets:
\bel{Cbar}
{\mathcal{C}}^{\textrm{PSPA}}(\beta, q_{0}) =
\prod_{r>0} \frac{\nu_{r}^{2} + \Gamma(q_{0}) \,\nu_{r} + \Omega^{2}(q_{0})}
{\nu_{r}^{2} + \Gamma(q_{0}) \,\nu_{r} + \varpi^{2}(q_{0})} \ .
\end{equation}
As mentioned earlier, in comparison to (\ref{def-omnu}) there is
only one (pair of) mode(s). The relation between the local
nucleonic frequency $\Omega(q_{0})$ and the local collective
frequency $\varpi(q_{0})$ is given by (\ref{freqcoll}). It is
worth stressing that the infinite product (\ref{Cbar}) is
convergent. To guarantee this important feature, it suffices to
have the same coefficients for local inertia and damping in the
numerator and the denominator \cite{wit.wat}.

Plugging (\ref{T0loc}), (\ref{ZS-min}) and (\ref{ZS-barr}) into
(\ref{-ImFrate}) we obtain the following expression for the PSPA
decay rate of the system under consideration:
\bel{PSPArate} R^{\textrm{PSPA}} = \frac{|\varpi_{b}|}{2\pi} \
\left( \sqrt{1 + \eta_{b}^{2}} - \eta_{b} \right) \
\sqrt{\frac{C_{\mathcal{F}}(q_{a})}{|C_{\mathcal{F}}(q_{b})|}} \
\frac{e^{-\beta{\mathcal{F}}^{\textrm{SPA}}(\beta, q_{b})}}
{e^{-\beta{\mathcal{F}}^{\textrm{SPA}}(\beta, q_{a})}} \times
\frac{{\mathcal{C}}^{\textrm{PSPA}}(\beta, q_{b})}{{\mathcal{C}}
^{\textrm{PSPA}}(\beta, q_{a})}
\end{equation}
Like in the sequel we have partly used indices "b" instead of an
argument $q_{b}$ to keep our notation short.
The two first factors, which in a sense represent dynamics, have
come in through the crossover temperature $T_0(q_{0})$ discussed
in section \ref{T0glob}, mind (\ref{T0loc}) in particular. Like in
the Caldeira-Leggett model the decay rate factorizes into a
classical part $R_{\textrm{class}}^{\textrm{PSPA}}$ and a quantum
correction factor $f_{\textrm{qm}}^{\textrm{PSPA}}$
\bel{rate-b}
R^{\textrm{PSPA}} =
R_{\textrm{class}}^{\textrm{PSPA}} \ \times \
f_{\textrm{qm}}^{\textrm{PSPA}}\,,
\end{equation}
with
\bel{rate-class}
R_{\textrm{class}}^{\textrm{PSPA}} =
\frac{|\varpi_{b}|}{2\pi} \,\sqrt{\frac{C_{\mathcal{F}}(q_{a})}
{|C_{\mathcal{F}}(q_{b})|}} \
\frac{e^{-\beta{\mathcal{F}}^{\textrm{SPA}}(\beta, q_{b})}}
{e^{-\beta{\mathcal{F}}^{\textrm{SPA}}(\beta, q_{a})}} \ \left(
\sqrt{1 + \eta_{b}^{2}} - \eta_{b} \right)
\end{equation} and \begin{equation} f_{\textrm{qm}}^{\textrm{PSPA}}=
\frac{{\mathcal{C}}^{\textrm{PSPA}}(\beta, q_{b})}{{\mathcal{C}}
^{\textrm{PSPA}}(\beta, q_{a})} \ ,
\end{equation}
respectively.

Let us discuss first the factor
$R_{\textrm{class}}^{\textrm{PSPA}}$ which survives the classical
limit, as no $\hbar$ is involved. Evidently, it contains the
common Arrhenius factor
\bel{Arrhenius} \exp \left( -\beta E_{\textrm{b}}\right) = \exp
\left( -\beta ({\mathcal{F}}^{\textrm{SPA}}(\beta, q_{b}) -
{\mathcal{F}}^{\textrm{SPA}}(\beta, q_{a})) \right) \ ,
\end{equation}
defined here by the difference $ E_{\textrm{b}}$ of the free
energy between barrier and potential minimum. The influence of
damping is given by the correction factor found first by Kramers
\cite{kr.40}, namely
\bel{KramFaktor}
f_{\textrm{K}} = \sqrt{1 + \eta_{b}^{2}} - \eta_{b} \ .
\end{equation}
It decreases monotonically with increasing $\eta_{b}$ and for
$\eta_{b} \gg 1$ behaves like $1 / 2\eta_{b}$. The remaining
factor can be made to become proportional to the attempt frequency
$\varpi_a$ at the minimum by writing
\bel{f-nuatt}
\frac{|\varpi_{b}|}{2\pi} \ \sqrt{\frac{C_{\mathcal{F}}(q_{a})}
{|C_{\mathcal{F}}(q_{b})|}} =
\frac{\varpi_{a}}{2\pi} \ \sqrt{\frac{M_{a}}{M_{b}}}
\ \sqrt{\frac{|C^{\textrm{coll}}(q_{b})|}{C^{\textrm{coll}}(q_{a})}}
\ \sqrt{\frac{C_{\mathcal{F}} (q_{a})}{|C_{\mathcal{F}}(q_{b})|}} \ .
\end{equation}
Here, use has been made of the relation between the frequency and
inertia of the local mode and the associated stiffness:
\bel{defstiff}
C^{\textrm{coll}}(q_{0}) = M(q_{0}) \,\varpi^{2}(q_{0})
\end{equation}
Putting all factors together the classical rate
may be written as
\bel{rate-class-2}
R_{\textrm{class}}^{\textrm{PSPA}} =
R_{\textrm{K}} \ f_{\textrm{class}}^{\textrm{PSPA}} \ .
\end{equation}
It contains Kramers' original form
\bel{rate-kram}
R_{\textrm{K}} = \frac{\varpi_{a}}{2\pi} \, e^{-\beta E_{\textrm{b}}}
\ \left( \sqrt{1 + \eta_{b}^{2}} - \eta_{b} \right)
\end{equation}
as the first factor. In addition there is a another correction
factor,
\bel{f-PSPA}
f_{\textrm{class}}^{\textrm{PSPA}} =
\sqrt{\frac{M_{a}}{M_{b}}} \ \sqrt{\frac{|C^{\textrm{coll}}(q_{b})|}
{C^{\textrm{coll}}(q_{a})}} \ \sqrt{\frac{C_{\mathcal{F}}(q_{a})}
{|C_{\mathcal{F}}(q_{b})|}}
\end{equation}
not present in the derivations based on the Caldeira-Leggett
model. The reasons for that are obvious. First of all, in this
model the inertia in the collective mode is simply put equal to a
constant which renders the first factor on the right of
(\ref{f-PSPA}) equal to unity. Second, the dynamical stiffness
(\ref{defstiff}) is forced to be identical to the one of the
phenomenologically introduced collective potential, $\partial^2
V(q)/\partial q^2$ in \cite{ca.le.83, weiss, in.98}.
This is achieved by working with a Hamiltonian in which
from the beginning the collective part is renormalized by the term
$\chi(0) q^2 / 2$ which in the linear response approach is induced
by the static influence of the coupling. In this way only the
dynamical part of the induced force appears, which in the end may
lead to Ohmic friction. In our approach, where {\em all transport
properties} of the collective dynamics are generated from the two
body interaction, such manipulations are not meaningful. In
certain limits it is possible, however, to simplify the
$f_{\textrm{class}}^{\textrm{PSPA}}$ of (\ref{f-PSPA}). For slow
collective motion, sometimes referred to as the zero frequency
limit, the local stiffness (\ref{defstiff}) of collective motion
may be shown to be represented by that of the free energy
(\ref{CFT-def}) (see e.g.\cite{hofrep}) such that one simply has
\bel{f-PSPA-Z}
f_{\textrm{class}}^{\textrm{PSPA}} \approx
\sqrt{\frac{M_{a}}{M_{b}}} \ .
\end{equation}
For a derivation of
this factor based on a generalized version of Kramers' equation
and picture of the decay we like to refer to \cite{hh.iv.ru.ya.01}.

Next we turn to the {\em quantum corrections} to the classical
rate (\ref{rate-class}), which in this approach is given by the
ratio of the PSPA corrections ${\mathcal{C}}^{\textrm{PSPA}}
(\beta, q_{0})$ of (\ref{Cbar}) evaluated at the barrier and the
minimum. With the help of the local nucleonic and collective
frequencies $\Omega(q_{0})$ and $\varpi(q_{0})$ it writes
\bel{fqm-PSPA}
f_{\textrm{qm}}^{\textrm{PSPA}} =
\prod_{r>0} \frac{\nu_{r}^{2} + \Gamma_{b} \,\nu_{r} + \Omega^{2}_{b}}
{\nu_{r}^{2} + \Gamma_{b} \,\nu_{r} + \varpi^{2}_{b}} \,:
\,\prod_{r>0} \frac{\nu_{r}^{2} + \Gamma_{a} \,\nu_{r} + \Omega_{a}}
{\nu_{r}^{2} + \Gamma_{a} \,\nu_{r} + \varpi^{2}_{a}}
\end{equation}
The nice feature about this structure is that it converges for
{\em all} conceivable values of the transport coefficients as long
as $T > T_{0}$. The reason simply is that it is the ratio of two
convergent products of type (\ref{Cbar}). As the alert reader may
guess a simple generalization of the quantum correction factor of
the Caldeira-Leggett model (see e.g. \cite{weiss, in.98}) to
coordinate-dependent coefficients  like
\bel{fqm-LC}
f_{\textrm{qm}}^{\textrm{LC}} \longrightarrow
\prod_{r>0} \frac{\nu_{r}^{2} + \Gamma_{a} \,\nu_{r} +
\varpi^{2}_{a}}{\nu_{r}^{2} + \Gamma_{b} \,\nu_{r} +
\varpi^{2}_{b}}
\end{equation}
(see e.g. \cite{fr.ti.92}) may (for Ohmic damping where $\Gamma$
does not fall off for large frequencies) lead to problems of convergence.
Indeed, for $\Gamma_a \neq \Gamma_b$ the infinite product either converges to
zero or diverges depending which one of the $\Gamma$'s is larger
\cite{wit.wat}. In \cite{hh.in.th.93} the form (\ref{fqm-LC}) has
been derived on the basis of a quantum transport equation. In
\cite{hh.iv.99} this factor has been evaluated for
microscopically calculated transport coefficients. To circumvent
the convergence problem in (\ref{fqm-LC}) the individual
$\Gamma$'s had been replaced by the arithmetic mean value
$2 \bar{\Gamma} = \Gamma_{a} + \Gamma_{b}$.

The local frequency of the collective motion $\varpi$ is real at
the minimum and purely imaginary at the barrier, whereas the
frequency of the nucleonic motion $\Omega$ is real everywhere. The
denominator of the first term in (\ref{fqm-PSPA}) vanishes as $T$
approaches $T_{0}$, corresponding to the definition of the
crossover temperature in section \ref{T0glob} (see (\ref{T0nur})),
but all other factors are strictly positive. For this reason,  at
$T_{0}$ the quantum correction factor
$f_{\textrm{qm}}^{\textrm{PSPA}}$ diverges to plus infinity, a
feature well known from the Caldeira-Leggett model
\cite{gr.we.ha.84}.

In the limit of very high temperatures $T \gg \hbar\Omega $, where
$\hbar\Omega$ represents the typical nucleonic excitation,
$f_{\textrm{qm}}^{\textrm{PSPA}}$ strictly converges to unity. For
nuclear fission collective motion is expected to be slow in the
sense \cite{hh.iv.ru.ya.01} of having
\bel{slowmot}
1 \,\textrm{MeV} \approx \hbar\varpi \ll \hbar\Omega \approx
\frac{41 \,\textrm{MeV}}{A^{1/3}} \approx 6 \,\textrm{MeV}
\end{equation}
One may expect quantum effects in collective motion to disappear
already for $T \gg \hbar\varpi $. Indeed, the quantum correction
factor will be close to unity already for $T / \hbar\Omega = {\cal
O}(1)$. At least this can be shown for the two factors of
(\ref{fqm-PSPA}). Divide all numerators and denominators by
$\nu_{r}^{2}$ and neglect $(\varpi / \nu_{r})^{2} \ll 1$ in the
resulting denominators. This is justified simply because the
condition $T / \hbar\Omega \approx {\cal O}(1)$ implies $\Omega /
\nu_{r} \approx 1 / (2\pi r)$ (mind (\ref{Matsubara})). Remain the
terms which involve $\Gamma / \nu_{r}$. They can be neglected if
$T \gg 1 / 2\pi \cdot \hbar\Gamma$ or $T / \hbar\varpi \gg \eta /
\pi$. Microscopic computations of the transport coefficients
show this condition to be fulfilled, although $\eta$ itself
increases with $T$; see figure 3 of \cite{hh.iv.ru.ya.01} or
figure 5.2.10 of \cite{hofrep}.

\section{Conclusion}
\label{concl}

We have been able to demonstrate how the PSPA can be extended to
treat the decay of damped meta-stable systems. In this first step
a simple schematic two body interaction has been taken and the
nucleonic excitations have been assumed to be concentrated in one
Lorentzian peak around a certain mean value. Generalizations to
more general systems should not cause too many problems. So far we
concentrated on the quantum corrections to thermal hopping which
take place above the critical temperature $T_0$
\cite{gr.we.ha.84, hh.in.th.93}. At this
temperature the common semi-classical treatment of functional
integrals breaks down, simply because for unstable modes the
Gaussian integrals diverge for smaller temperatures. So far this
latter feature also limited the applications of the PSPA to bound
systems \cite{la.pu.bo.bro.90, pu.bo.bro.91, at.al.97}.
There is hope that this deficiency can be overcome
in very much the same way as it was possible for dissipative
tunneling \cite{gr.we.ha.84, an.gr.95b}.
Work in this direction is under way \cite{ru.an.01}.

There are several advantages of the method presented here, both
over the usual approach to dissipative tunneling within the
Cladeira-Leggett model \cite{weiss, in.98}, as well as with respect to the
Locally Harmonic Apprtoximation (LHA) \cite{hofrep} to quantum
transport. Different to the Caldeira-Leggett model, all transport
properties derive from the two body interaction of the many body
system. No phenomenological assumptions have to be made for any
transport coefficient. The effects of the two body interaction are
treated on a fully self-consistent level, largely because the
collective variables can be introduced globally by way of the
Hubbard-Stratonovich transformation. For the LHA, on the other
hand, and on a quantum level this is possible only locally
\cite{hofrep}. This method, however, is more flexible with respect
to the thermal properties. There, one needs not rely on the concept
of a fixed temperature, an assumption which is questionable for
isolated systems.

\vspace{2cm}
{\bf Acknowledgments} \\
The authors would like to thank J.~Ankerhold and F.~Ivanyuk for interesting
discussions and helpful suggestions.


\begin{thebibliography}{99}
\bibitem{nat-adapt} Our notation relates to that of
\cite{at.al.97} in the following way:
$\hat{H}_{0} \leftrightarrow K$, $\hat{F} \leftrightarrow V$,
$k < 0 \leftrightarrow -\chi < 0$, $q \leftrightarrow \chi\sigma$,
$z \leftrightarrow \zeta_{0}$, ${\cal C} \leftrightarrow \zeta_{0}'$,
$\hbar \leftrightarrow 1$.
\end{thebibliography}
\end{document}